\documentclass[twocolumn]{pasj00}

\begin{document}
\SetRunningHead{Kashikawa et al.}{SDF Imaging Data}
\Received{2004/07/05}
\Accepted{2004/09/16}

\title{The Subaru Deep Field: The Optical Imaging Data\footnotemark[1]}



%
 \author{%


Nobunari \textsc{Kashikawa}	    \altaffilmark{2,3}
Kazuhiro \textsc{Shimasaku}        \altaffilmark{4}
Naoki \textsc{Yasuda}           \altaffilmark{5}
Masaru \textsc{Ajiki}            \altaffilmark{6}\\
Masayuki \textsc{Akiyama}          \altaffilmark{7}
Hiroyasu \textsc{Ando}             \altaffilmark{2}
Kentaro \textsc{Aoki}             \altaffilmark{7}
Mamoru \textsc{Doi}              \altaffilmark{8}
Shinobu S. \textsc{Fujita}        \altaffilmark{6}\\
Hisanori \textsc{Furusawa}         \altaffilmark{7}
Tomoki \textsc{Hayashino}        \altaffilmark{9}
Fumihide \textsc{Iwamuro}          \altaffilmark{10}
Masanori \textsc{Iye}              \altaffilmark{2,3}\\
Hiroshi \textsc{Karoji}           \altaffilmark{7}
Naoto \textsc{Kobayashi}        \altaffilmark{8}
Keiichi \textsc{Kodaira}          \altaffilmark{3}
Tadayuki \textsc{Kodama}           \altaffilmark{11}\\
Yutaka \textsc{Komiyama}         \altaffilmark{7}
Yuichi \textsc{Matsuda}          \altaffilmark{9, 2}
Satoshi \textsc{Miyazaki}         \altaffilmark{7}
Yoshihiko \textsc{Mizumoto}         \altaffilmark{2}\\
Tomoki \textsc{Morokuma}         \altaffilmark{8}
Kentaro \textsc{Motohara}         \altaffilmark{8}
Takashi \textsc{Murayama}         \altaffilmark{6}
Tohru \textsc{Nagao}            \altaffilmark{6, 12}\\
Kyoji \textsc{Nariai}           \altaffilmark{13}
Kouji \textsc{Ohta}             \altaffilmark{10}
Sadanori \textsc{Okamura}          \altaffilmark{4, 14}
Masami \textsc{Ouchi}            \altaffilmark{2}
Toshiyuki \textsc{Sasaki}           \altaffilmark{7}\\
Yasunori \textsc{Sato}             \altaffilmark{2}
Kazuhiro \textsc{Sekiguchi}        \altaffilmark{7}
Yasunori \textsc{Shioya}           \altaffilmark{6}
Hajime \textsc{Tamura}           \altaffilmark{9}\\
Yoshiaki \textsc{Taniguchi}        \altaffilmark{6}
Masayuki \textsc{Umemura}          \altaffilmark{15} 
Toru \textsc{Yamada}           \altaffilmark{2} 
and
Makiko \textsc{Yoshida}          \altaffilmark{4}
}

\footnotetext[1]{Based on data collected at Subaru Telescope, which is operated by the National Astronomical Observatory of Japan.}
\altaffiltext{2}{Optical and Infrared Astronomy Division, National Astronomical Observatory, \\
Mitaka, Tokyo 181-8588, Japan}
\altaffiltext{3}{The Graduate University for Advanced Studies (SOKENDAI), \\
Shonan Village, Hayama, Kanagawa 240-0193, Japan}
%
\altaffiltext{4}{Department of Astronomy, Graduate School of Science, University of Tokyo, Tokyo 113-0033, Japan}
\altaffiltext{5}{Institute for Cosmic Ray Research, University of Tokyo, Kashiwa 277-8582, Japan}
\altaffiltext{6}{Astronomical Institute, Graduate School of Science, Tohoku University, Aramaki, Aoba, \\
Sendai 980-8578, Japan}
\altaffiltext{7}{Subaru Telescope, National Astronomical Observatory of Japan, \\
650 N. A'ohoku Place, Hilo, HI 96720, USA}
\altaffiltext{8}{Institute of Astronomy, Graduate School of Science, University of Tokyo, \\
2-21-1 Osawa, Mitaka, Tokyo 181-0015, Japan}
\altaffiltext{9}{Research Center for Neutrino Science, Graduate School of Science, Tohoku University, \\ Aramaki, Aoba, Sendai 980-8578, Japan}
\altaffiltext{10}{Department of Astronomy, Graduate School of Science, Kyoto University, \\
Kitashirakawa, Sakyo, Kyoto 606-8502, Japan}
\altaffiltext{11}{Theoretical Astrophysics Division, National Astronomical Observatory, \\
Mitaka, Tokyo 181-8588, Japan}
\altaffiltext{12}{INAF -- Osservatorio Astrofisico di Arcetri, Largo Enrico Fermi 5, 50125 Firenze, Italy}
\altaffiltext{13}{Department of Physics, Meisei University, 2-1-1 Hodokubo, Hino, Tokyo 191-8506, Japan}
\altaffiltext{14}{Research Center for the Early Universe, Graduate School of Science, University of Tokyo, \\ Tokyo 113-0033, Japan}
\altaffiltext{15}{Center for Computational Physics, University of Tsukuba, \\
1-1-1 Tennodai, Tsukuba 305-8571, Japan}

%

\maketitle

\begin{abstract}

The Subaru Deep Field (SDF) project is a program of Subaru Observatory to carry out a deep galaxy survey over a blank field as large as $34' \times 27'$.
The program consists of very deep multi-band optical imaging, near infrared imaging for smaller portions of the field and follow-up optical spectroscopy.
Major scientific goals of the project are to construct large samples of Lyman-break galaxies at $z \simeq 4-5$ and Lyman alpha emitters at $z \simeq 5.7$ and $6.6$, and to make detailed studies these very high-redshift galaxy populations.

In this paper, we describe the optical imaging observations and data reduction, presenting mosaicked images and object catalogs in seven bandpasses.
The optical imaging was made through five broad-band filters, $B$, $V$, $R$, $i'$, $z'$, and two narrow-band filters, NB816 ($\lambda_c=8150$\AA) and NB921 ($\lambda_c=9196$\AA) with almost $10$ hours long integrations for each band.
The limiting magnitudes measured at $3 \sigma$ on a $2''$ aperture are $B=28.45$, $V=27.74$, $R=27.80$, $i'=27.43$, $z'=26.62$, ${\rm NB816}=26.63$, and ${\rm NB921}=26.54$ 
in the AB system.
The object catalog constructed for each of the seven bands contains more than $10^5$ objects. 
The galaxy number counts corrected for detection incompleteness and star count contribution are found to be consistent with previous results in the literature. 
The mosaicked images and catalogs of all the bands have been made open to the public on Oct. $1$, $2004$ on the SDF project website at http://soaps.naoj.org/sdf/.

\end{abstract}

\section{Introduction}

Deeper and wider-field observations are required to understand the universe in longer baseline of time and space.
A large galaxy survey is a simple and essential method in today's astronomy for the purpose.
However, an unprecedentedly large galaxy survey is made possible only with the combination of some distinctive resources, such as a large telescope, a unique and efficient instrument based on advanced technologies, and a large amount of telescope time created by the copperation of many people who appreciate the value of the survey.
In particular, such an extremely deep and wide galaxy surveys to collect a statistically robust number of high-$z$ galaxies beyond $z=3$ are definitely needed. 
There are a lot of high-$z$ galaxy surveys so far, for Lyman break galaxies (e.g., \cite{ste96,ste99}, \cite{iwa03}, \cite{gia04a}, \cite{ouc04a}) 
and Lyman $\alpha$ emitters (e.g., \cite{hu98}, \cite{aji02}, \cite{kod03}, \cite{rho03}, \cite{hu04}).
These surveys are sometimes not wide enough to overcome the effect of cosmic variance in deriving universal nature of these high-$z$ populations or sometimes not deep enough to catch a faint signal from these distant populations.

We carried out a large systematic survey searching for high-$z$ populations with Subaru Telescope, which we call the Subaru Deep Field (SDF) project.
Our major scientific goals of this project are to construct the largest samples of Lyman-break galaxies at $z\simeq 4-5$ and of Lyman $\alpha$ emitting galaxies at $z \simeq 5.7$ and $6.6$, and make detailed studies of these very high-redshift galaxy populations.
We are also carrying out a subsequent spectroscopy and near-infrared imaging for these targets.
In the deep optical imaging, we are going to take a couple of well-known approaches to search for these distant populations, one is to detect continuum break galaxies using deep multi broad-band imagings and the other is to isolate high-$z$ populations having only the Ly$\alpha$ emission feature by deep narrow-band imagings.
These are the conventional methods that have been taken in $4$m class telescopes and other $8$-$10$m class telescopes.
However, they become very powerful with Subaru and its unique wide-field camera. 

The SDF project was approved to be one of three Subaru Observatory Projects.
About $30$ nights were allocated for this SDF project during $2002$-$2004$ semesters.
The most unique point of the SDF is that the field is much wider than those of other similar deep surveys.
The Subaru prime-focus camera (Suprime-Cam; \cite{miy02}) has ten $2$k$\times4$k CCDs to cover a FOV as wide as $34'\times27'$.
The surveyed area is five times larger than a single GOODS field \citep{gia04b} and even $250$ times larger than the HDF.
With this very wide FOV, we can derive the statistical properties of distant galaxies such as luminosity function and angular correlation function from large samples which are robust against cosmic variance.
Samples of a large number of high-$z$ galaxies enable us to study different properties depending on various parameters, such as redshift, luminosity, and local density. 
Furthermore, the wide FOV gives us a high probability to discover rare objects such as the most distant galaxies.

In the SDF project, we attempted to take the deepest imaging data among those having a single Suprime-Cam's FOV.
We obtained almost $10$ hours long exposure time in each of $B$, $V$, $R$, $i'$ and $z'$ bands.
This extremely long integration yielded large catalogues that contain more than $100,000$ objects in each band.
In addition, we have also carried out $\sim10$ hours long integrated narrow-band imaging at $8160$\AA~and $9210$\AA, that correspond to dark night-sky windows free from OH emission lines.
These are similar wavelengths to those of other recent narrow-band deep surveys such as CADIS (e.g., \cite{mai03}), LALA (e.g., \cite{rho03}), and UH (e.g., \cite{hu04}) surveys.
These wavelengths corresponds to the redshifted Ly$\alpha$ emission at $z=5.7$ and $z=6.6$, respectively.
The initial discovery of two LAEs at $z=6.6$ on the SDF was reported in \citet{kod03}, and confirmation of an LAE at $z=6.33$ which was detected as an $i'$-dropout was presented in Nagao et al (2004 in preparation).

Though our survey is dedicated for high-$z$ populations, we can also sample very faint galaxies in the local universe. 
For example, the SDF enables studies of galaxies down to $M^*+3$ around $z=1$.
Also our deep narrow-band images contain many faint emission line galaxies in foreground. 

In this paper, we present the procedures to construct the optical imaging data set as well as catalogues 
that are released for public.
Observations are described in Section $3$, and data reduction processes are given in Section $4$.
We also show characteristics of the data such as $S/N$ distribution over the field, astrometric accuracies and completeness estimation (Section $6$). 
And finally in Section $7$, our completeness-corrected galaxy number counts are shown to be consistent with previous results.
Throughout this paper, magnitudes are in the AB system.

\section{The Field}

The Subaru Deep Field (SDF) is located near the North Galactic Pole, being centered on $(13^h 24^m 38.^s9, +27^{\circ} 29^{\prime} 25.^{\prime \prime}9)$ (J2000).
We selected this field by imposing the following three conditions: (1) absence of bright stars, bright galaxies, or nearby known cluster of galaxies within $\simeq 30'$ from the field center, (2) an extreme low Galactic extinction ($A_V=0.052$ at the center of the field; \cite{sch98}), and (3) a high altitude at meridional transit from Mauna-Kea and thus ensuring high observing efficiency (the maximum altitude of the SDF is $82^\circ$).

We conducted preliminary imaging and spectroscopic observations on the SDF during the commissioning runs of three instruments on the Subaru Telescope, Suprime-Cam (optical camera for the prime focus;  \cite{miy02}), CISCO (NIR camera;  \cite{mot02}), and FOCAS (optical camera and spectrograph for Cassegrain focus; \cite{kas02}) from $1999$ to $2001$.
We obtained very deep $J$ and $K'$ images ($J_{AB}=26.0$, $K'_{AB}=25.3$; \cite{mai01}) for a $2' \times 2'$ area near the field center.
Some studies based on the deep NIR data were reported in \citet{mai01, tot01a, tot01b, tot01c, nag02, kas03}.
As for the Suprime-Cam images in this preliminary survey, we obtained $B_{lim}=27.8$ ($3$hrs integration time), $V_{lim}=27.3$ ($2$ hrs), $R_{lim}=27.1$ ($1.5$ hrs), $i'_{lim}=26.9$ ($2$ hrs), $z'_{lim}=26.1$ ($1$ hr), which are merged into the new data to construct the final images.
Some results on this wide field optical data are presented in \citet{ouc03a, ouc04a, ouc04b, shi03, shi04}.

\section{Observations}

\begin{figure}
\FigureFile(160mm,200mm){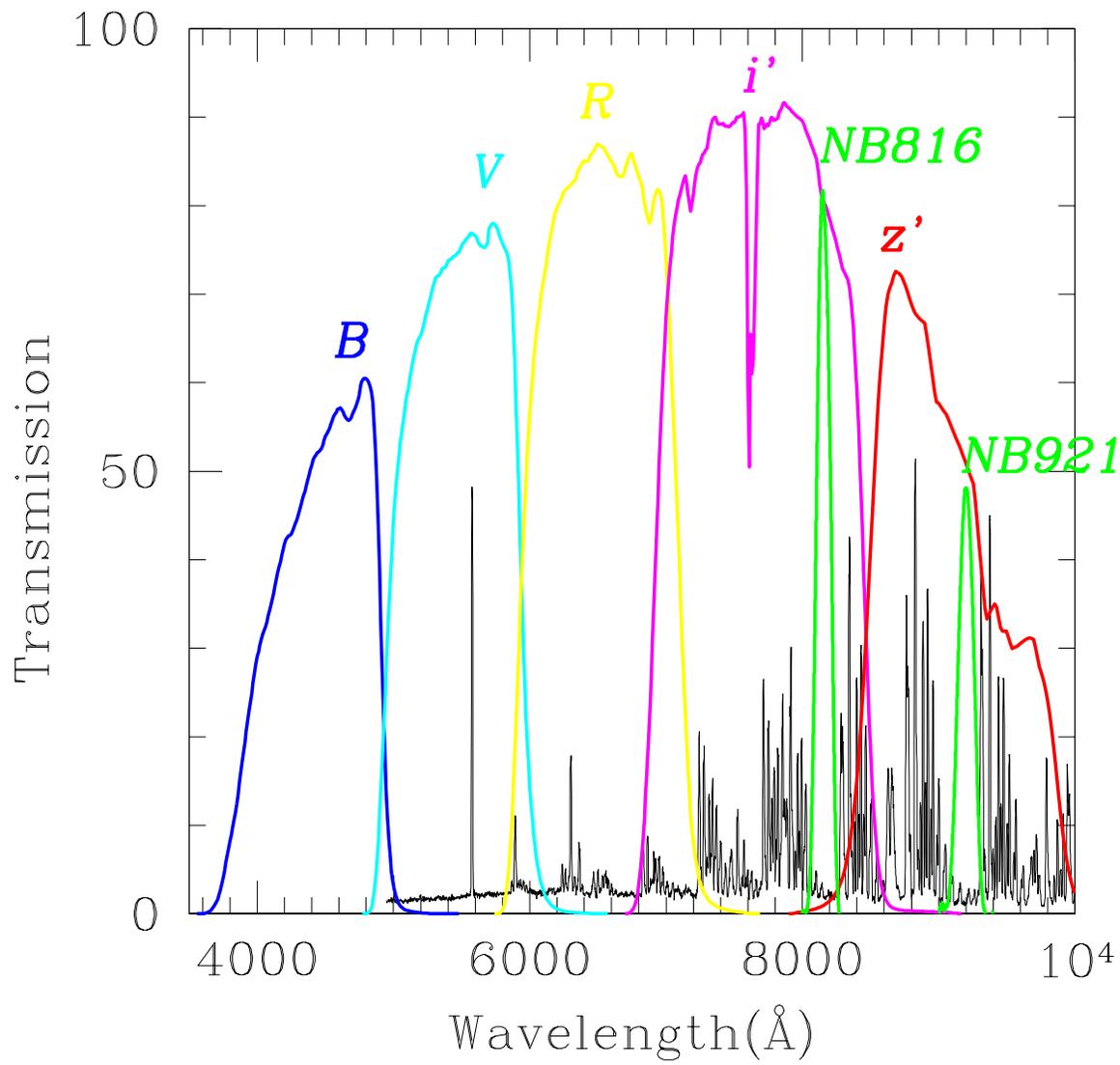}
\caption{The band response curves of SDF optical imaging. 
These are convolved with the CCD sensitivities, instrument and atmospheric transmissions.
Thin black line shows the OH night sky line spectrum.
\label{fig_filter}}
\end{figure}

Optical imaging was made with the prime-focus camera on the Subaru Telescope, Suprime-Cam.
Having ten $2$k $\times 4$k MIT/LL CCDs, Suprime-Cam covers a contiguous area of $34'\times27'$ with a pixel scale of $0.''202$ pixel$^{-1}$~\citep{miy02}.
The average gain of CCDs is $2.6e^-$/ADU.
The widths of the gaps between adjacent CCDs range over $79-84$ pixels (or $16\arcsec -17\arcsec$) in the north-south direction and $15-20$ pixels (or $3\arcsec -4\arcsec$) in the east-west direction.
We imaged the SDF in five standard broad-band filters, $B$, $V$, $R$, $i'$, and $z'$, and two narrow-band filters, 
NB816 ($\lambda_c=8150$\AA, FWHM$=120$\AA) and NB921 ($\lambda_c=9196$\AA, FWHM$=132$\AA) (see Figure~\ref{fig_filter}).
The two narrow-band filters, which have been carefully designed to avoid strong OH night sky emission lines, were used to search for Ly$\alpha$ emitting objects at $z\simeq 5.7$ and $z\simeq 6.6$, respectively.

The unit exposure time for each filter was determined so that the photon counts per pixel of the sky background reach an appropriate value. 
For the broadband filters, the typical unit exposure time is $900$ sec for $B$ and $240$ sec for $z'$, being shorter for redder filters because of the increase in the sky brightness with wavelength.
For the narrowband filters, the typical exposure time is $1800$ sec for both NB816 and NB921.

\begin{figure}
\FigureFile(160mm,200mm){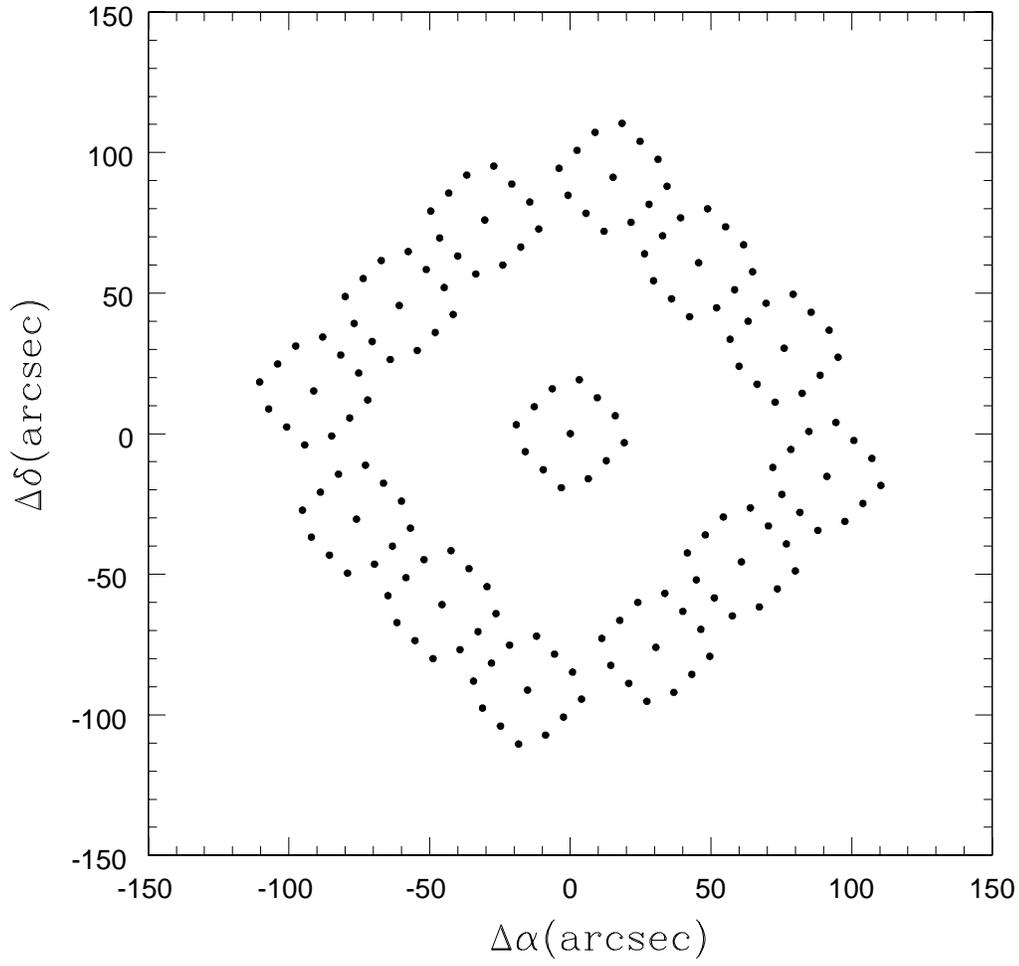}
\caption{The dithering pattern of observing pointings of SDF.
\label{fig_dith}}
\end{figure}

We adopted a common dithering pattern of pointings for all the bands. 
As shown in Figure~\ref{fig_dith}, a full cycle of dithering consists of $169$ pointings, which is divided into $13$ small {\lq}circles{\rq} of $13$ pointings.
In the actual imaging, the first $13$ pointings were made on a common position in the $13$ circles (the central point in each circle).
Similarly, the second $13$ pointings were made on another common position in the $13$ circles. 
The total number of pointings was less than $169$ (a full cycle of dithering) for any filters.
We devised the pattern of dithering, i.e., the distribution of $169$ pointings, 
taking account of the following conditions (some of which are in conflict with each other).
(1) The spread in pointings on the sky should be large enough to give sufficient overlaps between frames taken with adjacent CCDs. 
This condition is necessary, since the mosaicing of frames taken with adjacent CCDs is made using stars 
common to the frames.
(2) The spread in pointings on the sky should be small, in order to reduce areas of low $S/N$ ratios 
near the edges of the mosaiced image. 
(3) The mosaiced image should have sufficiently uniform $S/N$ ratios.
(4) In the course of dithering, no sky position should fall onto the same dead-pixel area more than once.
Each CCD has a number of long (but narrow) lines of dead pixels.
To meet condition (4), we adjusted the dithering positions in the horizontal ($X$) and vertical ($Y$) axes so that there are no duplicated $X$ values or $Y$ values in the $169$ positions.

As described below, the final images and catalogs are based only on the data with a seeing size of $<0.\arcsec 98$.
The fraction of the frames satisfying this seeing condition attains at least $70\%$ in exposure time for any of the filters.
The total exposure times, limiting magnitudes ($3\sigma$ on a $2''$ aperture), and observing dates are summarized in Table~\ref{tab_obs}.
Table~\ref{tab_obs} presents only the data good enough to be used to make final, mosaiced images. 
Low-quality data taken either in poor seeing (worse than $0."98$) or in very low transparency are not listed in the table.
All the raw data, including those with low quality, are available at SMOKA (Subaru Mitaka Okayama Kiso Archive) 
Science Archive\footnotemark[*].

\footnotetext[*]{http://smoka.nao.ac.jp/}

In constructing the final mosaiced images in the $B$, $V$, $i'$, and $z'$ bandpasses, we have co-added the data with $1-3$ hrs integration times taken during the commissioning observations of Suprime-Cam in $2001$.
The $i^\prime$-band data were taken separately over several observing runs because a high-$z$ supernova search project (Yasuda et al. in preparation) in this field requires intermittent observations in this band.
This repeated observation also enables the search for time-variable objects.

In the $B$, $V$, and $R$ observations, one of the three standard photometric standard fields, SA104, SA107, and SA110 \citep{lan92}, was observed for flux calibration, when the nights were photometric.
Similarly, in the $i'$, $z'$, NB816, and NB921 imaging, at least one of the five spectrophotometric standard stars, HZ-21, HZ-44, GD153, P177D, and P330E, was observed.

\section{Data Reduction}

We reduce all the data with good quality using a pipeline software package \citep{ouc03b} whose core programs were taken from IRAF and from software for mosaic-CCD data reduction developed by \citet{yag02}. 
The package includes bias subtraction, flat-fielding, correction for the image distortion due to the prime-focus optics, PSF matching (by Gaussian smoothing), sky subtraction, and mosaicing.
Bias subtraction and flat fielding were processed in the same manner as for the conventional single chip CCD. 
The flat frame was made from the median frame of a number of normalized object frames.
The distortion correction was performed with $4$-th order polynomial transformation~\citep{miy02} with $\sim0.5$ pix rms residuals.
The PSF matching was made for all frames so that the final mosaiced images in the seven bandpasses have a common PSF FWHM value of $0.\arcsec98$ through smoothing the images with adequate Gaussian kernels.
The global sky background was determined from bilinear interpolation of the typical sky counts of $64 \times 64$ pix ($12\arcsec.8 \times 12\arcsec.8$) meshes.
In mosaicing, first, the relative positions and relative throughputs between frames taken with different CCDs and at different exposures are calculated using stars common to adjacent frames.
Then, all the frames are merged into a single, large image by adjusting relative positions and stacking with weights according to the relative throughput of frames.
The geometry differences among seven band images were corrected with IRAF's {\tt geomap} and {\tt geotran} tasks.
About $1000$ objects were selected evenly over the image using SExtractor, and then spatial transformation law was determined from the iteration of $3$rd or higher order polynomial fitting with $3\sigma$ clipping unless the rms residuals were smaller than $0.1$pix.
%
%
The final mosaiced images for the seven bandpass are available at: http://soaps.naoj.org/sdf/.
The size of the final images is $8900\times 11000$ pixels (corresponding to $30' \times 37'$) after removal of very low-$S/N$ regions near the edges of the images.
In the future, we plan to make another set of SDF image data which consist of the images with good seeing sizes only in spite of shorter net integration times.

\subsection{Object Detection and Photometry}

We perform object detection and photometry by running SExtractor version 2.1.6 \citep{ber96} on the images.
The object detection is made for all seven images (five broadbands and two narrowbands) independently. 
For all objects detected in a given bandpass, their magnitudes and several parameters in the other six bandpasses are also measured at exactly the same positions as in the detection-band image, using the {\lq}double image mode{\rq} of SExtractor.
We detected objects that have $5$ connected pixels above $2\sigma$ of the sky background rms noise and made photometric measurements at $2\sigma$ level in all the catalogs.
Overlapping objects are deblended into distinct objects if the contrast between integrated intensity associated with each object is $\ge 0.005$.
Aperture photometry was done with $2\arcsec \phi$ and $3\arcsec \phi$ apertures.
Our input parameters on running SExtracter and the full list of the output parameters contained in the catalogs can also be browsed at: http://soaps.naoj.org/sdf/

\subsection{Photometric Calibration}

We calculate the photometric zero points of the images from the photometry of standard stars with a $10\arcsec \phi$ aperture, taking account of the differences in airmass between the standard-star frames and the fiducial SDF frames.
%
%
We check the photometric zero points using the stellar colors calculated from the spectrophotometric atlas of 175 Galactic stars given in \citet{gun83}. 
Since the FOV of Suprime-Cam is large, more than $100$ bright (but not saturated) stars with $i'\lesssim 23$ are commonly detected in mosaiced images of the seven bands.
We compare the location of the stellar sequence in various two-color planes between our data and Gunn \& Striker's atlas, finding that the colors of stars in our data are in agreement with those of Gunn \& Striker's stars within $\simeq 0.05$ mag.
Small offsets up to $\simeq 0.05$ mag seen in some combinations of bandpasses may indicate that the nights were not strictly photometric when the standard stars were taken for those bandpasses.
We correct the zero points of all the bandpasses by adding a small constant (at most $\simeq \pm 0.05$ mag) to the original value, so that the observed colors of stellar objects match completely the synthetic colors of \authorcite{gun83}'s stars.
In any case, the errors in the photometric-zero points of our final images would be less than $0.05$ mag.

\subsection{Limiting Magnitudes}

We measure the $3\sigma$ limiting magnitudes of the images, which are defined as the $3\sigma$ of the sky noise on a $2\arcsec \phi$ diameter.
For each image, we measure sky counts in a number of $2\arcsec \phi$ apertures which are located at randomly selected positions over the image. 
Then, we draw a histogram of the sky counts, and fit a Gaussian function to the histogram to obtain the $1\sigma$ noise.
When we fit a Gaussian function, we do not use the positive tail of the histogram which is affected by objects.
The limiting magnitudes thus estimated are presented in Table \ref{tab_obs}.
Note that we did not simply calculate the $3\sigma$ limiting magnitudes by scaling the $1\sigma$ noise in one pixel to a $2\arcsec $-diameter area, 
since such a simple scaling gives a very optimistic limiting magnitude as discussed in \citet{fur02} and \citet{lab03}.

\subsection{Astrometry}

The coordinates of the final, mosaiced images agree very well with the tangential coordinates, {\it i.e.}, coordinates obtained by projecting the sky onto a plane.  
Our mosaicing procedure using a large number of stars common to adjacent unit frames ensures very small random errors in the relative position (and relative throughput) over the entire mosaiced images; the relative errors are typically less than $1$ pixel (or $0.\arcsec 2$).

We then make absolute astrometric calibration as follows. 
We calibrate the coordinates (World Coordinate System) of the images using the objects common to catalogs based on our data and the 2MASS All-Sky Catalog of Point Sources (2MASS-PSC) \citep{cut03}. 
Calibrations are done for each band separately.  
The positional accuracies of 2MASS-PSC sources are as good as $70$-$80$mas over most of the magnitude range and $200$mas even at the faintest magnitudes of $K=16$. 
We use $385$ 2MASS-PSC sources which are located in the SDF area.  
They are uniformly distributed over the image. 
On the other hand, we have selected un-saturated SDF objects which have over $50$ connected pixels above $10\sigma$ of the sky background. 
Then we took cross-correlation between the positions of SDF objects and 2MASS-PSC sources. 
We find $277$-$290$ objects are matched between two catalogs after rejection of some obvious disagreements due to object blending.  
The faintest magnitudes of matched objects are $23.03$, $21.85$, $20.88$, $19.89$, $18.75$, $18.89$ and $18.66$ for $B$, $V$, $R$, $i'$, $z'$, $NB816$ and $NB921$ band, respectively. 
Calculations of WCS have been done using IRAF's {\tt images.imcoords.ccmap} task.

\begin{figure}
\FigureFile(160mm,200mm){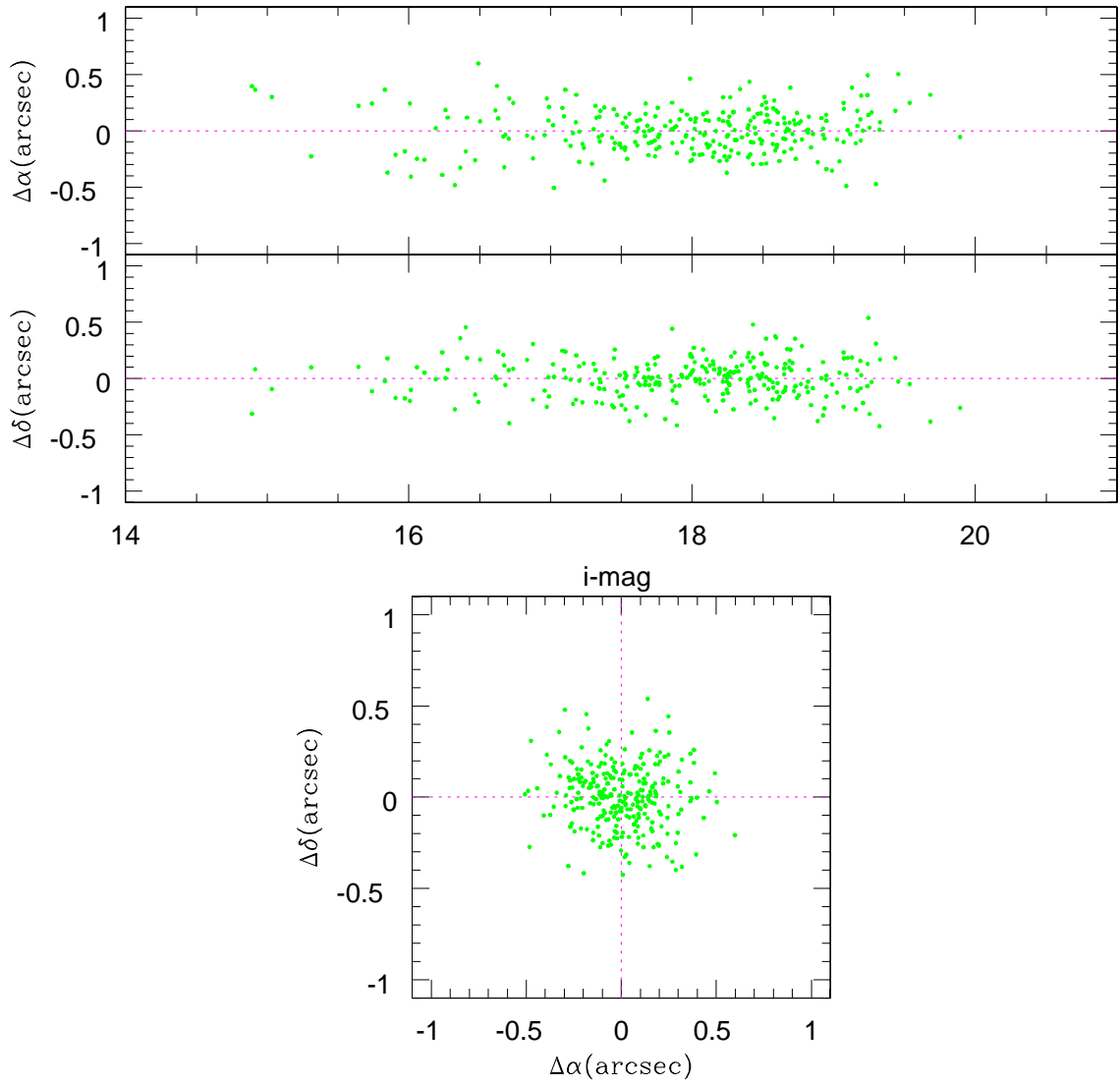}
\caption{Astrometry accuracy of SDF in the case of $i'$-band.
$\Delta\alpha$ and $\Delta\delta$ are calibration residuals between SDF objects and 2MASS-PSC sources in the directions of right ascension and declination, respectively.
\label{fig_ast}}
\end{figure}

The fitting procedure made a small residuals as shown in Figure~\ref{fig_ast}. 
The absolute position accuracy was found to be $0.224$, $0.258$, $0.266$, $0.259$, $0.214$, $0.222$ and $0.209$ arcsec, and rms errors are $0.154$, $0.198$, $0.205$, $0.196$, $0.152$, $0.160$ and $0.144$ arcsec in right ascension, $0.163$, $0.167$, $0.171$, $0.171$, $0.150$, $0.154$ and $0.152$ arcsec in declination for $B$, $V$, $R$, $i'$, $z'$, $NB816$ and $NB921$ band, respectively.
Figure~\ref{fig_ast} also shows that there is no systematic dependence on the magnitude and no significant offsets either in right ascension and declination.

\section{Catalogs}

We make seven catalog set for each detection bandpass.
For instance, the $B$-band detected catalog set contains seven measured catalogs which list the position, magnitudes, and several parameters concerning size and shape, of objects measured in the $B$-band image, as well as the magnitudes in the other six bandpasses measured for the same position and aperture as in the $B$-band image.

\begin{figure}
\FigureFile(160mm,200mm){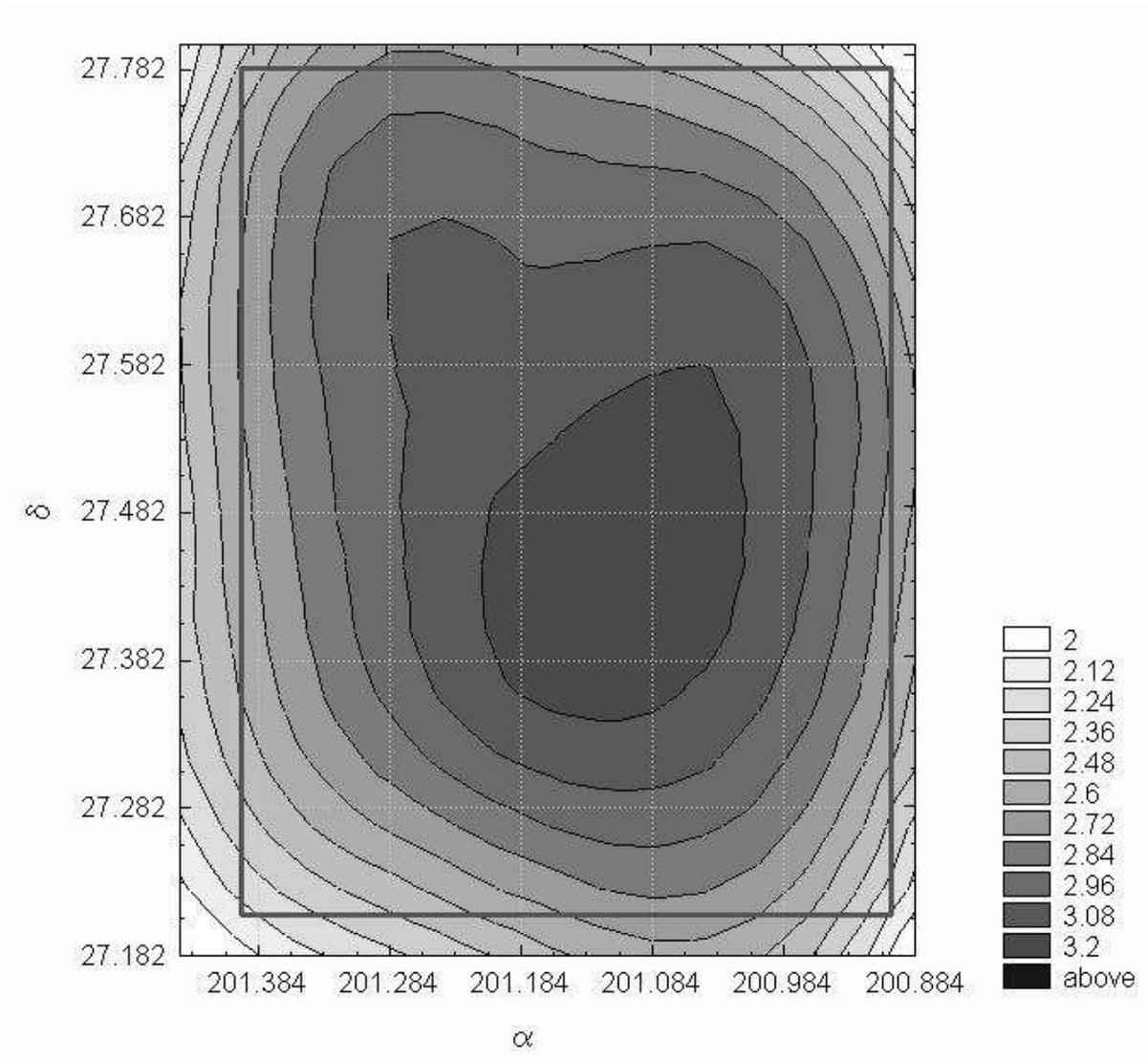}
\caption{The $S/N$ map of SDF in the case of $i'$-band. 
Thick (red) line shows the trimmed area defined in the ^^ ^^ masked region" file.
\label{fig_sn}}
\end{figure}

The efficiency and reliability of object detection and photometry are significantly lower than the average around very bright stars (due to bright haloes and saturation trails) and near the edges of the images (due to low $S/N$ ratios). 
We carefully defined these low-quality regions (^^ ^^ masked regions") which correspond to low-quality regions such as frame edges and the area around very bright stars with saturation trails.
The final catalogs are not masked, though it is better for most of cases to remove objects falling onto these regions.
Figure~\ref{fig_sn} plots the $S/N$ distribution map on the final image in the case of the $i'$-band.
The sky noise values are evaluated in the same manner as was described in Section $4.3$ for each of $355\times355$ pixel meshes with steps overlapping a half size of the mesh.
Then the $S/N$ values are calculated in the case of the objects with the limiting magnitude of $i'=27.43$.
Darker regions imply higher-$S/N$ regions.
Figure~\ref{fig_sn} demonstrates that our dithering pattern gives a fairly uniform distribution of $S/N$ ratios over the region.
The almost axisymmetric distribution reflects the vignetting pattern of prime-focus optics and the slightly lower $S/N$ region near the edge in the left-hand side would be due to the auto-guider probe which sometimes shadowed the FOV.
The lower left-hand corner with considerably lower $S/N$ ratios reflects the low quantum efficiency of the CCD which covered this region; the quantum efficiency of this CCD is about $60\%$ those of the others due to an incomplete AR coating~\citep{miy02}.
Similar $S/N$ maps have been obtained for the other filters.

The number of objects contained in the final catalogs are $214,733$ ($B$), $198,466$ ($V$), $209,477$ ($R$), $198,581$ ($i'$), $187,647$ ($z'$), $195,369$ (NB816), and $197,363$ (NB921).
If we remove objects that are positiond in the masked regions, the numbers of objects are reduced to  $162,611$ ($B$), $150,991$ ($V$), $160,129$ ($R$), $153,030$ ($i'$), $126,332$ ($z'$), $133,242$ (NB816), and $119,493$ (NB921).
The catalogs as well as the file which defines the masked regions are available at: http://soaps.naoj.org/sdf/.

\section{Completeness}

In this section, we describe the simulation carried out to estimate the completeness of our catalogs.
For that purpose, we distribute artificial objects on the original images.
To characterize an ensemble of the artificial galaxies, we have to assume the distributions of magnitude, surface brightness, luminosity profile.
These parameters strongly influence on the estimate of sample completeness.
Though the completeness of galaxy sample should be determined as a function of both magnitude and surface brightness, our sample is so deep that it is not sure that the assumptions for these faint galaxies are valid.
Therefore, we avoid an inaccurate estimate for extended objects whose completeness significantly depends on surface brightness, and alternatively, we present more secure evaluation of the sample completeness for point sources.
The extended objects would have brighter complete limits than those we show here.

\begin{figure}
\FigureFile(160mm,200mm){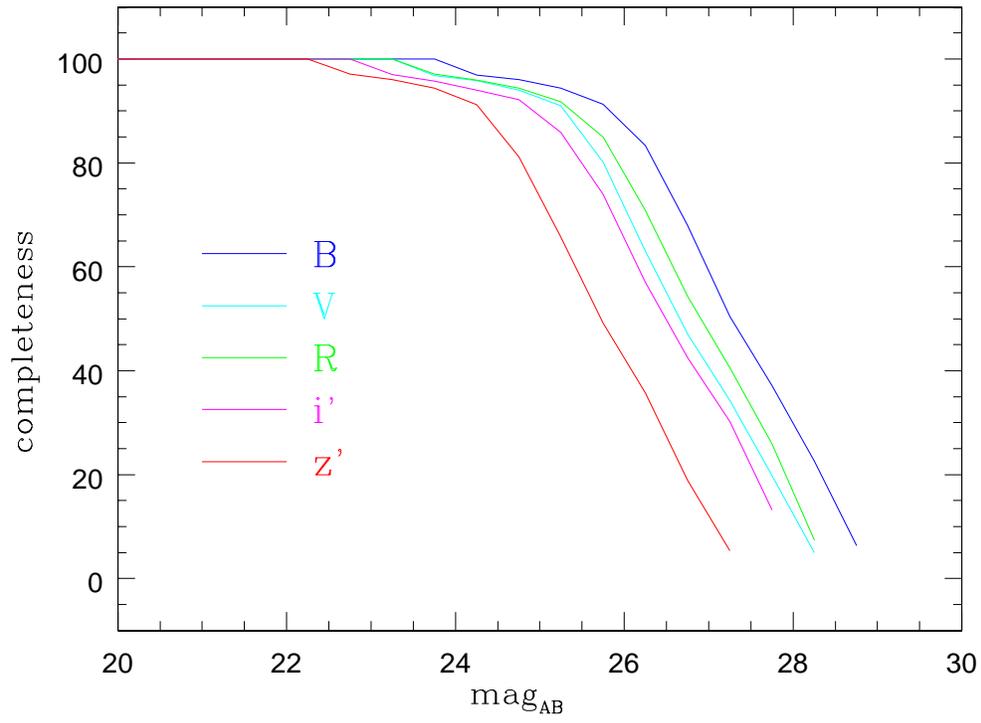}
\caption{The completess for our final SDF sample as a function of magnitudes.
\label{fig_compl}}
\end{figure}

We used IRAF tasks of {\tt starlist} and {\tt mkobjects} to make artificial stars on the original images.
Artificial stars have the same FWHM ($0.98$arcsec) as the real images, and are randomly distributed on the original images.
This random layout sometimes places artificial objects on real objects, which causes a heavy blending and incorrect photometry.
Thus, we neglect an artificial object if the distance from it to the nearest real object is less than three times the FWHM of the real object.
This rejection procedure means that we evaluate the object detection efficiency on pure sky background regions, 
and thus may give slightly optimistic estimation for completeness because object blending actually reduces the efficiency.
We extract artificial stellar objects with SExtractor using the same parameter set as adopted in the catalog construction (Section $5$).
Then we measure the fraction in number of detected objects to the input sources.
We generate $50,000$ artificial objects for a realization and repeat it four times to obtain a statistically robust number ($\sim 10,000$) of objects for each $0.5$ magnitude bin.
Figure~\ref{fig_compl} presents the resulting completeness estimates for each broadband as a function of total magnitude ({\tt MAG$\_$AUTO} magnitudes from SExtractor).

\section{Galaxy Counts}

In this section, we compare the galaxy number counts in the SDF with those given in the literature for each broad band.
When deriving the galaxy counts from our catalogs, we remove bright ($<24$ mag) Galactic stars from the catalogs, by imposing the following criteria: 
\begin{eqnarray}
I_{\rm star}\geq 0.99 & ({\rm mag} \leq 20), \\
{\rm log(FWHM)} \leq (24-{\rm mag})/150 & (20<{\rm mag} \leq 24). 
\end{eqnarray}
where $I_{\rm star}$ is the stellarity index, an output from SExtractor, FWHM is in arcsec, and mag is the total magnitude.
We do not make star/galaxy separation for objects fainter than $24$ mag, since in those magnitudes a significant fraction of galaxies are too small to securely discriminate from stars under the seeing size of our images ($0.\arcsec98$).
However, the surface density of stars is expected to be much lower than that of galaxies in those faint magnitudes.
We then correct the raw galaxy counts for detection completeness.
On the other hand, our galaxy catalogs can include false spurious objects.
We estimate the contamination by spurious objects by running SExtractor on negative images in the same manner as for the original images. 
We find that the contamination is as low as $2\%$ even in the faintest magnitude bins of our number 
counts. 
Thus, we do not correct our counts for this small contamination by spurious objects.

Figure $6$(a-e) plots the galaxy counts derived from our data together with those taken from the literature.
The filled circles and open circles correspond, respectively, to the raw counts and the corrected counts.
Since the raw counts include Galactic stars, they are higher than the corrected counts at the bright end of the plots.
These counts are also presented in Table~\ref{tab_nm}.
The solid lines shown at the bright end for the $B$, $V$, and $R$ plots are predicted star counts toward the SDF based on the stellar population synthesis model for the Galaxy by \citet{rob03}.
We have confirmed that the star counts calculated from our data ($<24$ mag) are consistent with the predicted counts in all three bandpasses.
Note that $i'$- and $z'$-band predictions are not available.

%

The galaxy number counts in the SDF data are reliable up to $\simeq 28$ mag for $B$, $\simeq 27.5$ mag for $V$, $R$, and $i'$, and $\simeq 26.5$ mag for $z'$, beyond which the completeness drops below $30\%$.
At these faint-end magnitudes, the counts reach $\simeq 3 \times 10^5$ galaxies per $0.5$ mag per square degree 
for all five bandpasses.
Comparison between our counts and those given in the literature shows broad agreement over the entire overlapping magnitude range for all bandpasses.
The SDF galaxy counts cover as wide as $8$ magnitudes from $20$ mag up to $\sim 28$ mag. 


\section{Summary}

The paper describes an overview of the SDF project and the procedures to complete the optical imaging data and catalogues.
We have carried out a deep optical imaging in seven filters, $B$, $V$, $R$, $i'$, $z'$, NB816, and NB921 over the $34' \times 27'$ in the SDF.
The limiting magnitudes measured at $3 \sigma$ on a $2''$ aperture are $B=28.45$, $V=27.74$, $R=27.80$, $i'=27.43$, $z'=26.62$, ${\rm NB816}=26.63$, and ${\rm NB921}=26.54$ in the AB system.
Object catalogs constructed from these extremely deep data contain more than $10^5$ objects in each bandpass.
The galaxy number counts corrected for detection completeness are found to be consistent with previous results in the literature. 

Scientific results based on the data will be shown in the forthcoming papers.
We hope that the public released SDF data will help in advances of our understanding for the high-$z$ universe as well as for a variety of astronomical topics.

\vskip 0.5cm

We deeply appreciate the devoted technical and managemental support of the Subaru Telescope staff to this long-term project.
The observing time for this project was committed to all the Subaru Telescope builders.
We thank the referee for helpful comments that improved the manuscript.
The research is supported by the Japan Society for the Promotion of Science through Grant-in-Aid for Scientific Research 16740118.

\clearpage

\begin{table}
\begin{center}
\caption{Summary of Optical Imaging Data for SDF.} \label{tab_obs}
\begin{tabular}{ccccccl}
\hline\hline
band & exp.time  & seeing size & $m_{lim}$\footnotemark[$*$] & $m_{0}$\footnotemark[$\dagger$]  &  $N$\footnotemark[$\ddagger$] & Date of Observations \\
 & (min.) & (arcsec) & (mag) & (mag/count) \\
\hline
$B$        & $595$ & $0.98$ & $28.45$ & $34.780$ & $162,611$ & 2003, Mar. 31, Apr. 2 \\
$V$        & $340$ & $0.98$ & $27.74$ & $34.584$ & $150,991$ & 2003, Apr. 30 \\ 
$R$        & $600$ & $0.98$ & $27.80$ & $34.315$ & $160,129$ & 2003, Mar. 30, Apr. 1/2/25/29/30 \\
$i^\prime$ & $801$ & $0.98$ & $27.43$ & $34.253$ & $153,030$ & 2002, Apr. 11/14, May 6, \\
           &       &        &         &          &           & 2003, Mar. 31, Apr. 2/24/25/29/30 \\
$z^\prime$ & $504$ & $0.98$ & $26.62$ & $33.001$ & $126,332$ & 2002, Apr. 9/14, 2003, Mar. 7, Apr. 1/28 \\
$NB816$    & $600$ & $0.98$ & $26.63$ & $32.880$ & $133,242$ & 2003, Apr. 28/29/30 \\
$NB921$    & $899$ & $0.98$ & $26.54$ & $32.520$ & $119,493$ & 2002, Apr. 9/11/14, May 6, \\
           &       &        &         &          &           & 2003, Mar. 7/8, Apr. 24 \\
\hline
\multicolumn{4}{@{}l@{}}{\hbox to 0pt{\parbox{150mm}{
\footnotemark[$*$] The $3\sigma$ limiting magnitudes in the AB system within $2^{\prime \prime}$ diameter aperture.
  \par\noindent
\footnotemark[$\dagger$] The magnitude zero point.
  \par\noindent
\footnotemark[$\ddagger$] The number of detected objects out of the masked regions.
  }\hss}}
\end{tabular}
\end{center}
\end{table}

\begin{table}
\begin{center}
\caption{Number counts on SDF.\footnotemark{$*$}} \label{tab_nm}
\begin{tabular}{cccccc}
\hline\hline
Magnitude & $B$  & $V$ & $R$ & $i'$  & $z'$ \\
\hline
18.250	& 2.076	1.882	& 2.246	2.169	& 2.350	2.169	& 2.607	2.280	& 2.642	2.234	\\			
18.750	& 2.155	1.155	& 2.386	2.076	& 2.597	2.269	& 2.739	2.394	& 2.802	2.523	\\		
19.250	& 2.394	1.792	& 2.597	2.222	& 2.695	2.291	& 2.848	2.504	& 2.885	2.592	\\			
19.750	& 2.368	1.909	& 2.674	2.301	& 2.868	2.597	& 2.955	2.678	& 3.100	2.919	\\			
20.250	& 2.477	2.155	& 2.805	2.570	& 2.990	2.812	& 3.149	3.000	& 3.282	3.185	\\			
20.750	& 2.682	2.394	& 3.015	2.848	& 3.175	3.055	& 3.360	3.258	& 3.486	3.399	\\			
21.250	& 2.848	2.651	& 3.124	3.019	& 3.361	3.279	& 3.529	3.472	& 3.644	3.595	\\			
21.750	& 3.056	2.941	& 3.322	3.233	& 3.544	3.477	& 3.697	3.651	& 3.847	3.808	\\			
22.250	& 3.265	3.195	& 3.537	3.479	& 3.743	3.698	& 3.882	3.848	& 3.959	3.927	\\			
22.750	& 3.535	3.480	& 3.747	3.711	& 3.918	3.891	& 4.011	3.986	& 4.113	4.101	\\			
23.250	& 3.774	3.747	& 3.976	3.955	& 4.087	4.068	& 4.177	4.181	& 4.261	4.261	\\			
23.750	& 4.048	4.034	& 4.194	4.192	& 4.290	4.293	& 4.354	4.367	& 4.417	4.430	\\			
24.250	& 4.285	4.299	& 4.410	4.429	& 4.480	4.498	& 4.516	4.549	& 4.559	4.599	\\			
24.750	& 4.483	4.501	& 4.590	4.617	& 4.649	4.674	& 4.661	4.698	& 4.694	4.785	\\			
25.250	& 4.652	4.677	& 4.739	4.780	& 4.770	4.807	& 4.781	4.846	& 4.790	4.972	\\			
25.750	& 4.779	4.819	& 4.846	4.942	& 4.871	4.942	& 4.873	5.007	& 4.863	5.171	\\			
26.250	& 4.882	4.961	& 4.932	5.133	& 4.940	5.089	& 4.932	5.179	& 4.885	5.332	\\			
26.750	& 4.952	5.119	& 4.973	5.300	& 4.981	5.246	& 4.942	5.329	&		\\			
27.250	& 4.989	5.287	& 4.953	5.418	& 4.966	5.357	&		&		\\			
27.750	& 4.973	5.404	&		&		&		&		\\			
\hline
\multicolumn{4}{@{}l@{}}{\hbox to 0pt{\parbox{130mm}{
\footnotemark[$*$] The first columns show the raw counts in log(/deg$^2$/$0.5$mag), and the second column show the counts corrected for completeness and star counts for each band.
  }\hss}}
\end{tabular}
\end{center}
\end{table}

\clearpage





\clearpage

\clearpage

\clearpage

\clearpage

\clearpage

\clearpage

\begin{figure}
\FigureFile(160mm,200mm){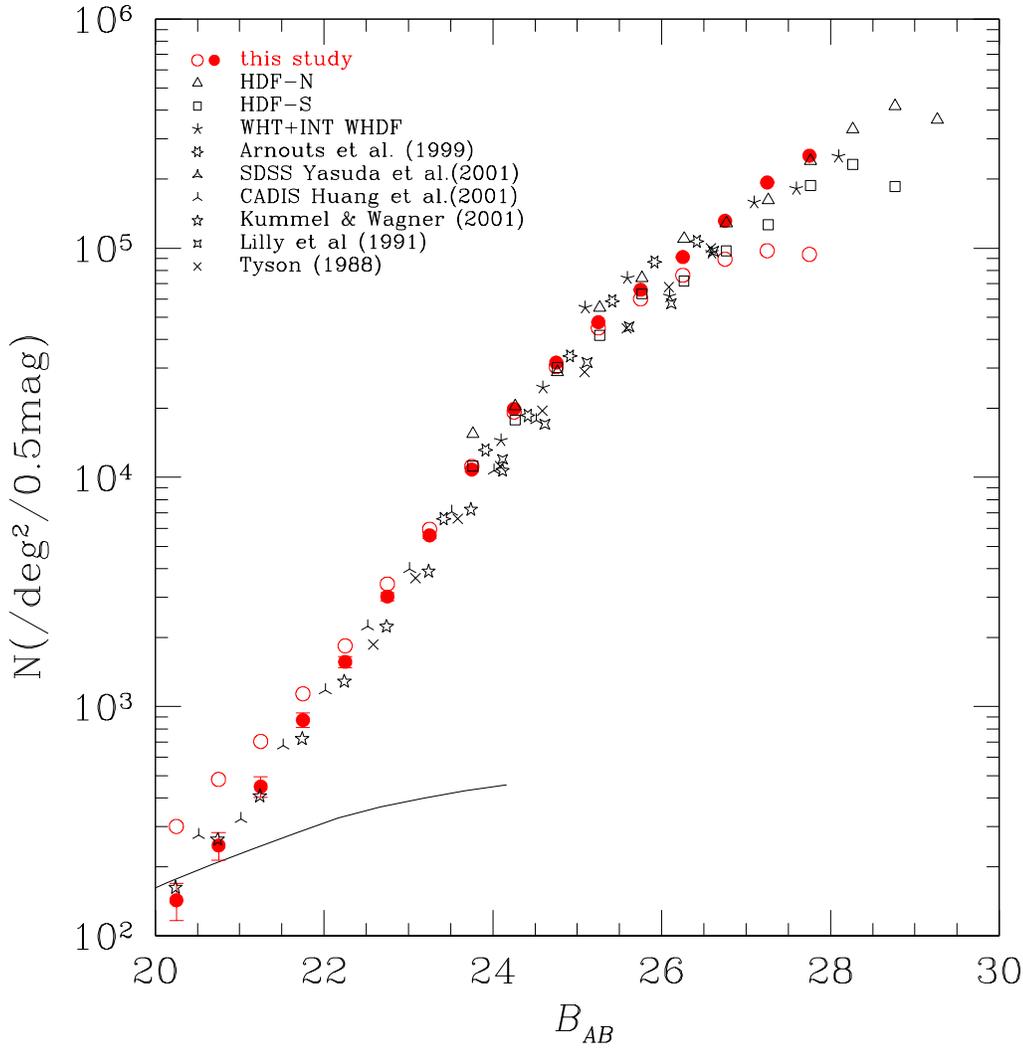}
\caption{The galaxy number counts in B-band.
Open (red) circles show the raw number counts from the SDF catalog, filled (red) circles show the galaxy number counts corrected with completeness and subtracted star counts, and other symbols from the literature.
Solid line shows the predicted star counts of \citet{rob03}. The error bars shown in the SDF counts are based on the simple Poisson errors.}
\label{fig_nmb}
\end{figure}
\clearpage

\begin{figure}
\FigureFile(160mm,200mm){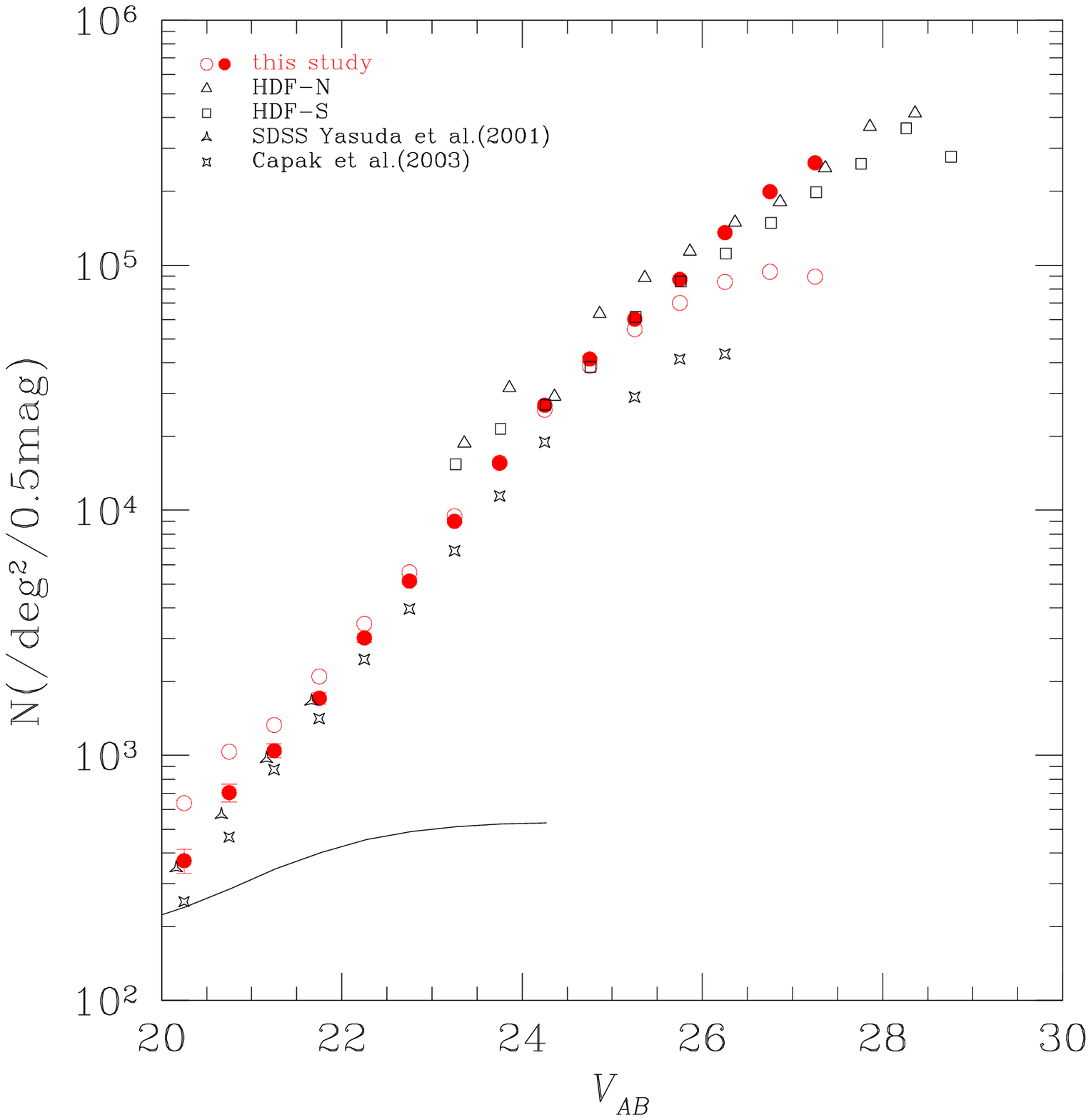}
\caption{Same as in Figure~\ref{fig_nmb} but in V-band.}
\label{fig_nmv}
\end{figure}
\clearpage

\begin{figure}
\FigureFile(160mm,200mm){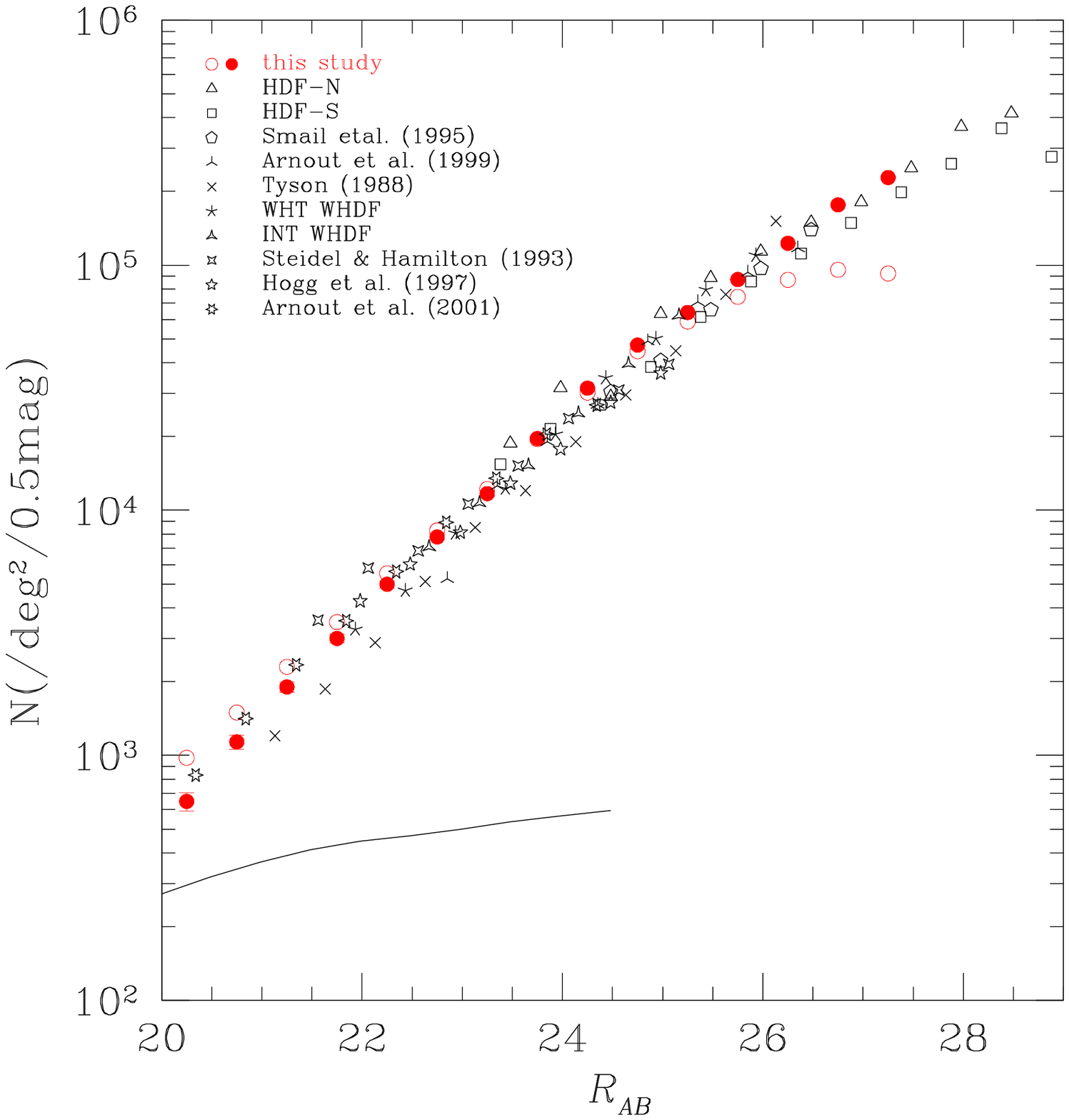}
\caption{Same as in Figure~\ref{fig_nmb} but in R-band.}
\label{fig_nmr}
\end{figure}
\clearpage

\begin{figure}
\FigureFile(160mm,220mm){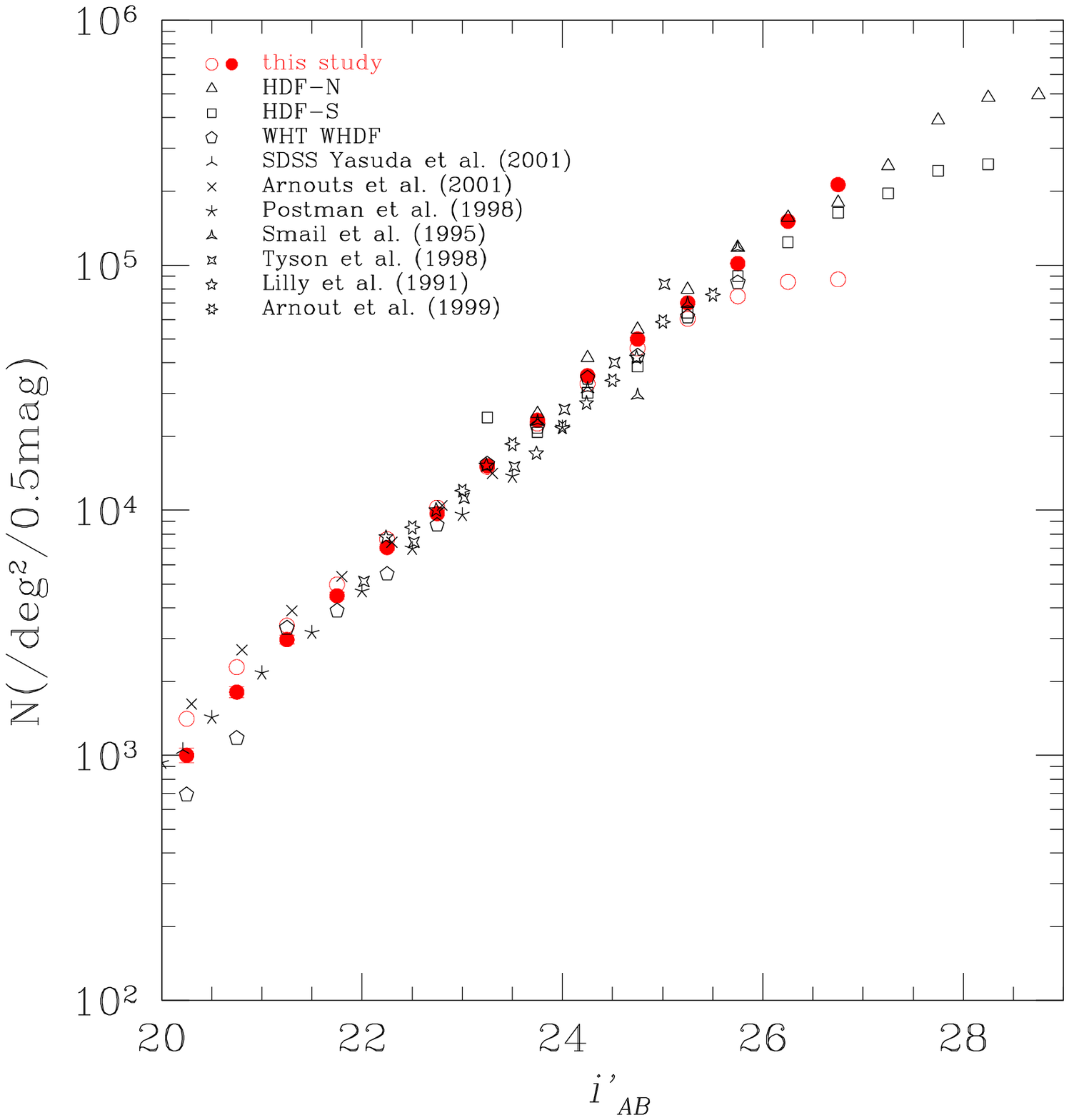}
\caption{Same as in Figure~\ref{fig_nmb} but in i'-band.}
\label{fig_nmi}
\end{figure}
\clearpage

\begin{figure}
\FigureFile(160mm,220mm){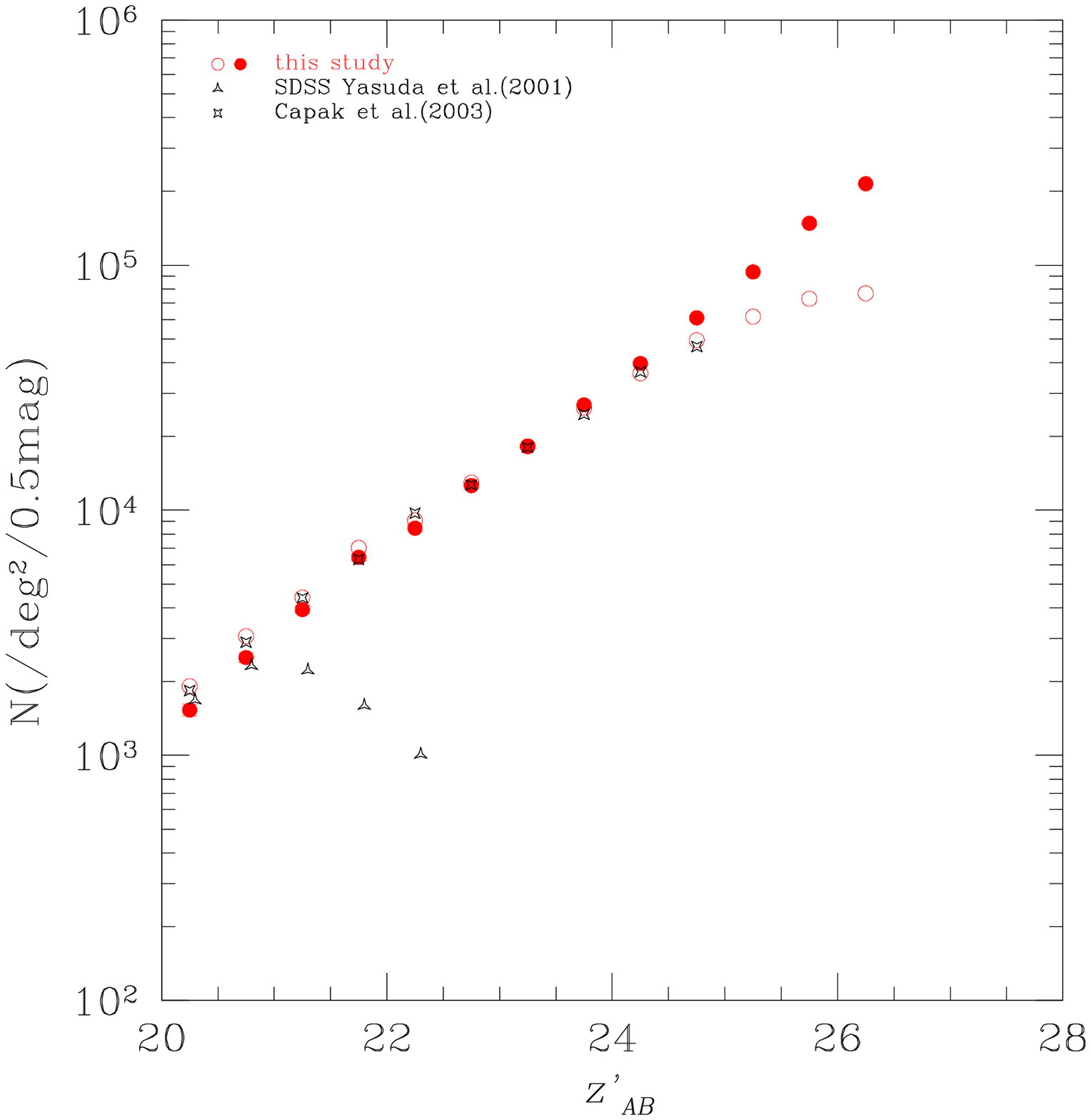}
\caption{Same as in Figure~\ref{fig_nmb} but in z'-band.}
\label{fig_nmz}
\end{figure}
\clearpage 

\end{document}